\documentclass{SolarPhysics}    
\usepackage[optionalrh]{spr-sola-addons}
\usepackage{epsfig}
\usepackage{url}
\usepackage{fixltx2e}
\usepackage[usenames]{color}

\newcommand{\ang}{\AA\ }
\newcommand{\gapprox}{\lower.4ex\hbox{$\;\buildrel >\over{\scriptstyle\sim}\;$}}
\newcommand{\lapprox}{\lower.4ex\hbox{$\;\buildrel <\over{\scriptstyle\sim}\;$}}

\def\ang{\AA}
\def\etal{{\it et al.,~}}
\def\eg{{\it e.g.,~}}
\def\ie{{\it i.e.,~}}

\def\aap  {{\sl Astron. Astrophys.}\ }   
\def\apj  {{\sl Astrophys. J.}\ }        
\def\apjl {{\sl Astrophys. J. Lett.}\ }  
\def\sp   {{\sl Solar Phys.}\ }          

\begin{document}
\begin{article}
\begin{opening}
\title{A Nonlinear Force-Free Magnetic Field Approximation 
	Suitable for Fast Forward-Fitting to Coronal Loops. 
	III. The Free Energy}

\author{Markus J. Aschwanden}
\runningauthor{Aschwanden}
\runningtitle{Free Magnetic Energy}

\institute{Solar and Astrophysics Laboratory,
	Lockheed Martin Advanced Technology Center, 
        Dept. ADBS, Bldg.252, 3251 Hanover St., Palo Alto, CA 94304, USA; 
        (e-mail: \url{aschwanden@lmsal.com})}

\date{Received ... ; Accepted ...}

\begin{abstract}
An analytical approximation of a nonlinear force-free magnetic field
(NLFFF) solution was developed in Paper I, while a numerical code 
that performs fast forward-fitting of this NLFFF approximation to 
a line-of-sight magnetogram and coronal 3D loops has been described 
and tested in Paper II. Here we 
calculate the free magnetic energy $E_{\rm free}=E_{\rm N}-E_{\rm P}$, \ie the
difference of the magnetic energies between the nonpotential field
and the potential field. A second method to estimate the free energy
is obtained from the mean misalignment angle change 
$\Delta\mu=\mu_{\rm P}-\mu_{\rm N}$ 
between the potential and nonpotential field, which scales as $E_{\rm free}/E_{\rm P}
\approx \tan^2{(\Delta\mu)}$. For four active regions observed with STEREO
in 2007 we find free energies in the range of $q_{\rm free}=(E_{\rm free}/E_{\rm P}) \approx
1\%-10\%$, with an uncertainty of less than $\pm 2\%$ between the two methods,
while the free energies obtained from 11 other NLFFF codes exhibit a larger
scatter of order $\approx\pm10\%$. We find also a correlation 
between the free magnetic energy and the GOES flux of the largest 
flare that occurred during the observing period, which can be 
quantified by an exponential relationship, $F_{\rm GOES} \propto 
\exp{(q_{\rm free}/0.015)}$, implying an exponentiation of the dissipated 
currents.
\end{abstract}


\end{opening}

\section{	Introduction 				}

The free magnetic energy is the maximum amount of energy that can be
released in an active region of the solar corona, such as during a solar 
flare, a filament eruption, or a coronal mass ejection (CME). Thus,
it is important to design reliable methods and tools that can calculate 
the amount of free energy in order to quantify the energy budget in 
a catastrophic energy release event, as well as for estimating upper limits
in forecasting individual events in real-time. Traditionally, the free energy is 
calculated by computing the nonpotential field ${\bf B}_{\rm N}({\bf x})$
with a numerical nonlinear force-free field (NLFFF) code and a potential field 
${\bf B}_{\rm P}({\bf x})$ for the same photospheric boundary data 
${\bf B}(x, y, z_{\rm phot})$, 
so that the difference of the magnetic field energy density integrated 
over a volume $V$ encompassing the active region of interest can be quantified as
\begin{equation}
	E_{\rm free} = {1 \over 8\pi} \left( \int {\bf B}_{\rm N}^2({\bf x)}\ dV 
	 	- \int {\bf B}_{\rm P}^2({\bf x)}\ dV \right) \ .
\end{equation}
This standard method may not necessarily reflect the correct amount of
maximum free energy released during a solar flare, since the magnetic field
in the photospheric boundary ${\bf B}(x, y, z_{\rm phot})$ may
change during a flare (\eg see measurements by Wang \etal 1994, 2002, 
2004, 2013; Wang, 1997, 2006; Wang and Liu, 2010). Another problem with NLFFF
codes using the photospheric vector field is the non-force-freeness
of the lower chromosphere (Metcalf \etal 1995; DeRosa \etal 2009), 
which however, can be ameliorated by preprocessing the magnetic
boundary data, using chromospheric 
field measurements (\eg Metcalf \etal 2005; Jing \etal 2010), 
or by a multigrid optimization that minimizes
a joint measure of the normalized Lorentz force and the divergence
of the magnetic field, as proposed by Wiegelmann (2004) and applied
by Jing \etal (2009). A quantitative comparison of NLFFF computation
methods, however, revealed a substantial scatter of free energies
in the order of $\approx\pm10\%$ (of the potential energy), depending on 
the numeric code, the boundary specifications, and the spatial resolution 
(Schrijver \etal 2006; DeRosa \etal 2009).
Alternatively, some studies show that the free
energy is better estimated by the minimum-energy state above the linear
force-free field with the same magnetic helicity (Woltjer, 1958;
R\'egnier and Priest, 2007). Since NLFFF calculations are rather
computing-intensive for forward-fitting tasks to coronal constraints,
which requires many iterations (for an overview and discussion of different
numerical methods see recent reviews by Aschwanden (2004) or 
Wiegelmann and Sakurai (2012)) or faster non-numerical methods are desirable.
Some proxy of the active region's free magnetic energy has been defined 
based on the twist and magnetic field orientation near the neutral line 
(\eg Falconer \etal 2006, 2011). 

Free magnetic energies have been calculated for a variety of solar phenomena,
for instance for the evolution of the free magnetic energy during flux 
emergence and cancellation, using NLFFF codes (\eg Fang \etal 2012), 
for the evolution of active regions (\eg Kusano \etal 2002),
for helmet-shaped streamer configurations (Choe and Cheng, 2002), or
for breakout CMEs, using MHD simulations (\eg DeVore and Antiochos, 2005). 
Theoretical studies quantify the evolution of free magnetic energy
for dipolar (R\'egnier, 2009) and quadrupolar magnetic configurations 
with a null-point, using different force-free models (R\'egnier, 2012). 

An analytical approximation of a nonlinear force-free magnetic field 
(NLFFF) solution was developed in Paper I (Aschwanden, 2012), while
a numerical code that performs fast forward-fitting of this NLFFF
approximation to coronal 3D loops and a line-of-sight magnetogram
has been described and tested in Paper II (Aschwanden and Malanushenko,
2012). In this Paper III of the series we are concerned with the
calculation of the free energy, generally defined by the
difference of the magnetic energies between the nonpotential field 
and the potential field (Equation (1)).
We calculate free energies from simulated data (from Paper II), from the
analytical NLFFF solution of Low and Lou (1990), from active region
NOAA 10930 during an X3.4 flare modeled by Schrijver \etal (2008)
and Malanushenko \etal (2012), and from stereoscopically
triangulated loops observed with STEREO (Aschwanden \etal 2012).  
Section 2 describes the analytical treatment, Section 3 the application
to simulated datasets and observations, Section 4 contains a
discussion, and Section 5 the conclusions.

\section{	Analytical Formulation  				}

\subsection{The NLFFF Approximation}

In Paper I (Aschwanden, 2012) we derived an analytical approximation of a
nonlinear force-free field (NLFFF) solution, which fulfills Maxwell's
divergence-free equation ($\nabla \cdot {\bf B} = 0$) and 
the force-free equation $(\nabla \times {\bf B}) = \alpha({\bf x}) {\bf B}$
with second-order accuracy (of the force-free parameter $\alpha$). 
The analytical approximation can be specified by a radial field
$B_r$ and an azimuthal field component $B_\varphi$,
\begin{equation}
        B_r(r, \theta) = B_j \left({d^2 \over r^2}\right)
        {1 \over (1 + b^2 r^2 \sin^2{\theta})} \ ,
\end{equation}
\begin{equation}
        B_\varphi(r, \theta) =
        B_j \left({d^2 \over r^2}\right)
        {b r \sin{\theta} \over (1 + b^2 r^2 \sin^2{\theta})} \ ,
\end{equation}
\begin{equation}
        B_\theta(r, \theta) \approx 0
        \ ,
\end{equation}
\begin{equation}
        \alpha(r, \theta) \approx {2 b \cos{\theta} \over
        (1 + b^2 r^2 \sin^2{\theta})}  \ ,
\end{equation}
where ($r, \varphi, \theta$) are the spherical coordinates of a
single magnetic field component ($B_j, x_j, y_j, z_j, b_j)$
with a unipolar magnetic charge $B_j$ that is buried at position
($x_j, y_j, z_j)$, has a depth $d=1-[x_j^2-y_j^2-z_j^2]^{1/2}$,
a vertical twist $\alpha=2 b_j$, 
and $r=[(x-x_j)^2+(y-y_j)^2+(z-z_j)^2]^{1/2}$
is the distance of an arbitrary coronal position $(x,y,z)$ to the
subphotospheric location $(x_j, y_j, z_j)$ of the buried magnetic charge.
The force-free parameter $\alpha$ is expressed in terms
of the parameter $b$, which quantifies the number $N_{\rm twist}$
of full twist turns over a (loop) length $L$,
\begin{equation}
        b = {2 \pi N_{\rm twist} \over L} .
\end{equation}
This analytical approximation is divergence-free and force-free
to second-order accuracy in the parameter $(b\ r \sin \theta)$,
which is approximately proportional to the force-free
parameter $\alpha$ as defined by Equation (5). 

A general magnetic field configuration can be composed by a 
superposition of $N_{\rm m}$ twisted magnetic field components,  
\begin{equation}
        {\bf B}_{\rm N}({\bf x}) = \sum_{j=1}^{N_{\rm m}} {\bf B}_j({\bf x}) \ ,
\end{equation}
which also fulfils the divergence-free and force-free condition
with second-order accuracy in $\alpha$ (or $b$). 

\subsection{The Free Magnetic Energy of a Single Twisted Component}

Let us calculate now the free magnetic energy $dE_{\rm free}({\bf x})$ 
at location $({\bf x})$ for
the field resulting from a single twisted (buried) magnetic charge
as defined by Equations (1) to (6). Since the radial $B_r$ and
azimuthal components $B_{\varphi}$ are always orthogonal to each
other (Figure 1), we can calculate the total nonpotential magnetic field 
strength $B_{\rm N}$ at every given point $(r, \varphi)$ simply from the sum 
of the squared components $B_r$ and $B_\varphi$,
\begin{equation}
	B_{\rm N} = \left( B_r^2 + B_\varphi^2 \right)^{1/2} 
          = B_j \left({d^2 \over r^2}\right)
        {1 \over \sqrt{(1 + b^2 r^2 \sin^2{\theta})}} \ ,
\end{equation}
while the field strength $B_{\rm P}$ of a potential field corresponds to the
radial component $B_r$ (of a single buried magnetic charge, Equation (2)),
\begin{equation}
        B_{\rm P} = B_r = B_j \left({d^2 \over r^2}\right)
                    {1 \over (1 + b^2 r^2 \sin^2{\theta})} \ ,
\end{equation}
and thus the free energy $dE_{\rm free}({\bf x})$ is just the magnetic energy 
associated with the azimuthal field component $B_{\varphi}$, with 
Equations (8) and (9),
\begin{equation}
	dE_{\rm free}{(\bf x}) = dE_{\rm N}({\bf x}) - dE_{\rm P}({\bf x}) 
		 = {1\over 8\pi} [B_r({\bf x})^2+B_\varphi({\bf x})^2] 
	         - {1\over 8\pi} B_r({\bf x})^2
	         = {1\over 8\pi} B_\varphi({\bf x})^2. 
\end{equation}
This definition of the free magnetic energy $dE_{\rm free}({\bf x})$ fulfills the
following conditions:

\begin{figure}
\centerline{\includegraphics[width=0.5\textwidth]{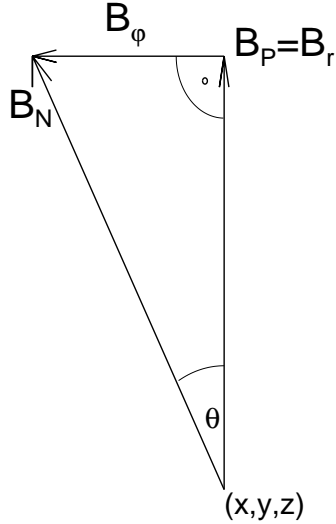}}
\caption{Diagram of the nonpotential field vector ${\bf B}_{\rm N}$,
which is composed of the two orthogonal components of the 
azimuthal field vector ${\bf B}_\varphi$ and the radial field
vector ${\bf B}_r$, subtending an angle $\theta$.}
\end{figure}

\begin{enumerate}
\item{\underbar{Positivity constraint:} The free energy is positive for 
	every nonpotential field at any location ${\bf x}$, 
	\ie $dE_{\rm free}({\bf x}) > 0$, since $B_\varphi^2({\bf x}) \ge 0$ 
	and $dE_{\rm free}({\bf x}) \propto B_\varphi^2({\bf x})$ 
	according to Equation (10).}
\item{\underbar{Additivity of energies and orthogonality of magnetic
	field components:}{\ } The nonpotential energy corresponds to 
	the sum of the potential energy and the free energy 
	$dE_{\rm N}({\bf x})=dE_{\rm P}({\bf x})+dE_{\rm free}({\bf x})$ 
	according to Equation (1) and Figure 1.
	Since the energies scale with the square of the magnetic
	field components, \ie $dE_{\rm N}({\bf x}) \propto B_{\rm N}({\bf x})^2$, 
	$dE_{\rm P}({\bf x}) \propto B_r({\bf x})^2$,
	and $dE_{\rm free}({\bf x}) \propto B_\varphi({\bf x})^2$, the Pythagoras theorem
	can be applied and it follows that $B_r$, $B_{\varphi}$, and $B_{\rm N}$ 
	form a rectangular triangle, with $B_{\varphi}$ perpendicular
	to the potential field component $B_{\rm P}=B_r$.}
\item{\underbar{Positive scaling with force-free parameter:} The free energy
	quantifies an excess of nonpotential magnetic field energy
	compared with the potential field energy, and thus should have a
	positive scaling with the force-free parameter $\alpha$. 
	Since the free energy scales proportional to the square of the 
	azimuthal magnetic field component (Equation (10)), we have a
	positive scaling, which is approximately $dE_{\rm free}({\bf x}) \propto b^2 
	\propto \alpha^2$ (Equations (3) and (5)).}
\item{\underbar{Potential field limit:} The free energy 
	vanishes asymptotically ($dE_{\rm free}({\bf x}) \mapsto 0$) with vanishing
	force-free parameter $|\alpha| \mapsto 0$ or 
	$|b| \mapsto 0$, with the potential field being the limit,
	$dE_{\rm N}({\bf x}) \mapsto dE_{\rm P}({\bf x})$.}
\item{\underbar{Finiteness of nonpotential energy:} The total nonpotential
	magnetic energy integrated over an arbitrary large height $r$
	converges to a finite value, $E(r \mapsto \infty)=E_{\rm max}$.
	We can prove the finiteness of the potential 
	energy for a single magnetic charge, which has a square-dependence of
	the magnetic field, $B(r) \propto r^{-2}$, yielding a 4th-power dependence
	of the magnetic energy $dE_{\rm P}(r) \propto B(r)^2 \propto r^{-4}$, and thus
	a 3th-power dependence for the integrated magnetic energy,
	$E_{\rm P,tot} \propto \int dE_{\rm P}(r) dr \propto r^{-3}$. For a finite
	amount of twist and a finite number of magnetic sources, it can be
	shown that the integral of the resulting nonpotential energy 
	is also finite.}
\end{enumerate}

\subsection{The Free Magnetic Energy of Multiple Twisted Components}

While the foregoing definition of the free energy is calculated for
a single twisted (buried) magnetic charge, how can it be generalized
for a superposition of an arbitrary number of magnetic charges, 
as defined in Equation (7)? The sum of the magnetic field contributions
from each buried magnetic charge component ${\bf B}_j({\bf x})$ add up to the 
nonpotential field vector ${\bf B}_{\rm N}({\bf x})$, which can be decomposed 
into two orthogonal components ${\bf B}_\parallel({\bf x})$ and 
${\bf B}_\perp({\bf x})$ in every point of space $({\bf x})$, 
\begin{equation}
        {\bf B}_{\rm N}({\bf x}) = \sum_{j=1}^{N_{\rm m}} {\bf B}_j({\bf x}) 
	= {\bf B}_\parallel({\bf x}) + {\bf B}_\perp({\bf x}) \ ,
\end{equation}
where the parallel component is aligned with the potential field
direction, ${\bf B}_\parallel \parallel {\bf B}_{\rm P}$, and the 
perpendicular component is orthogonal to the potential field direction, 
${\bf B}_\perp \perp {\bf B}_{\rm P}$. For a single (twisted) magnetic component
the parallel component ${\bf B}_\parallel$ is identical with the radial 
component ${\bf B}_r$, and the perpendicular component ${\bf B}_{\perp}$
is identical with the azimuthal component ${\bf B}_\varphi$ (Figure 1). 
The three magnetic field components $B_{\parallel}=B_{\rm P}, B_{\perp}$, and 
$B_{\rm N}$ are then associated each with one of the three energy components,
\begin{equation}
	\begin{array}{ll}
	dE_{\rm free}({\bf x})&=(1/8\pi) B_\perp^2({\bf x}) \\
	dE_{\rm P}({\bf x})   &=(1/8\pi) B_\parallel^2({\bf x}) \\
	dE_{\rm N}({\bf x})   &=(1/8\pi) [B_\parallel^2({\bf x})+B_\perp^2({\bf x})]
	\ .
	\end{array}
\end{equation}
Alternatively, the magnetic energy $dE_{\rm P}({\bf x)}$ of the potential field 
can be computed by using current-free magnetic field components 
($B_j, x_j, y_j, z_j, b_j=0$) straightforward with Equation (1), and the 
magnetic energy $dE_{\rm N}({\bf x})$ of the non-potential field with the current 
components ($B_j, x_j, y_j, z_j, b_j \neq 0$) with Equation (1) also, which 
yields the free energies $(E_{\rm free}=E_{\rm N}-E_{\rm P})$, after volume integration. 
Both methods are fitting the same line-of-sight component of the photospheric 
boundary $B_z(x,y,z_{\rm phot})$ given by the magnetogram, while
the non-potential magnetic field affects the transverse field components 
$B_x(x,y,z_{\rm phot})$ and $B_y(x,y,z_{\rm phot})$ that are not used as a boundary 
condition in our forward-fitting method. 

\begin{figure}
\centerline{\includegraphics[width=1.0\textwidth]{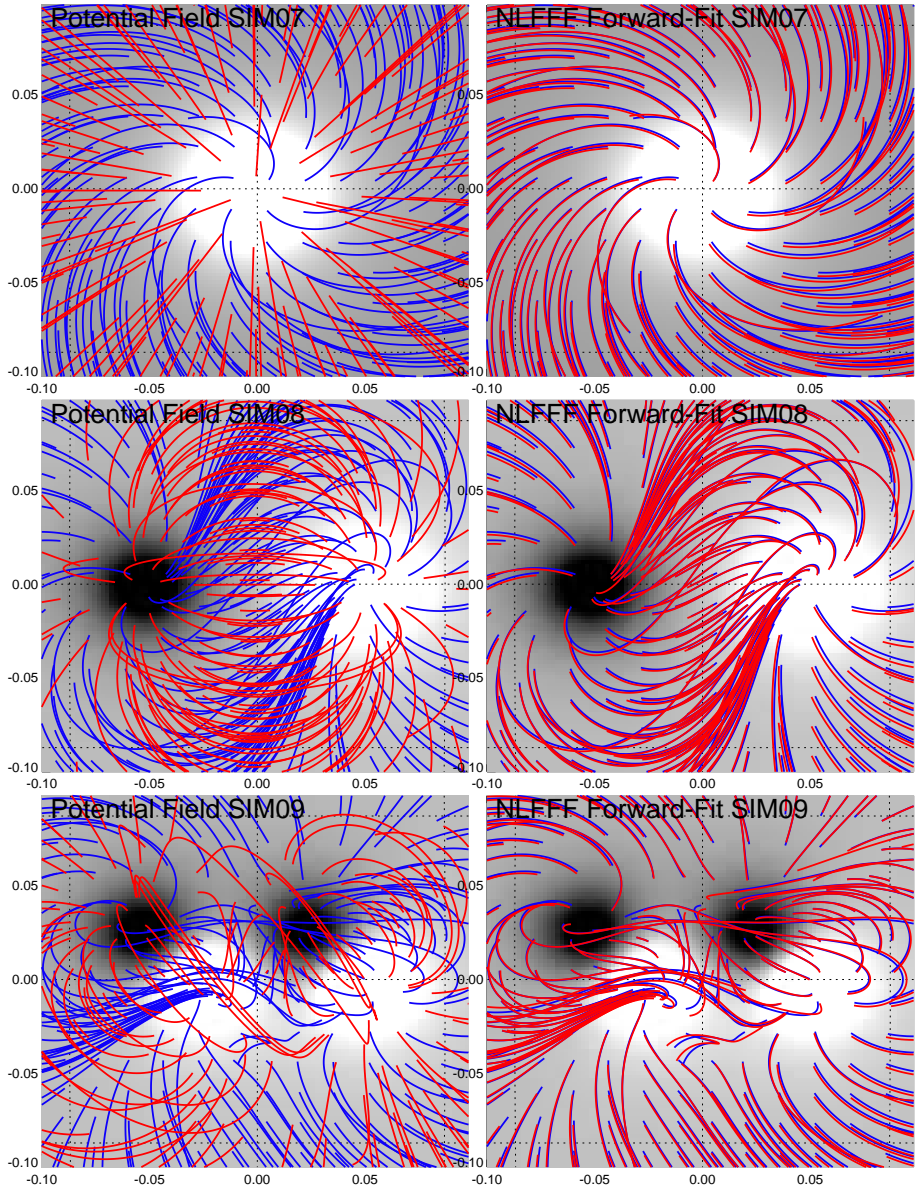}}
\caption{Forward-fitting of simulated nonpotential field data 
(cases N7-N9). Each panel shows the line-of-sight magnetogram
(grey), simulated coronal loops (the targets of the forward fit) (blue), 
and magnetic field lines of a theoretical model (red), either the 
potential field constrained by the line-of-sight magnetogram (left panels), 
or the forward fit of the NLFFF approximation (right panels).}
\end{figure}

\begin{figure}
\centerline{\includegraphics[width=1.0\textwidth]{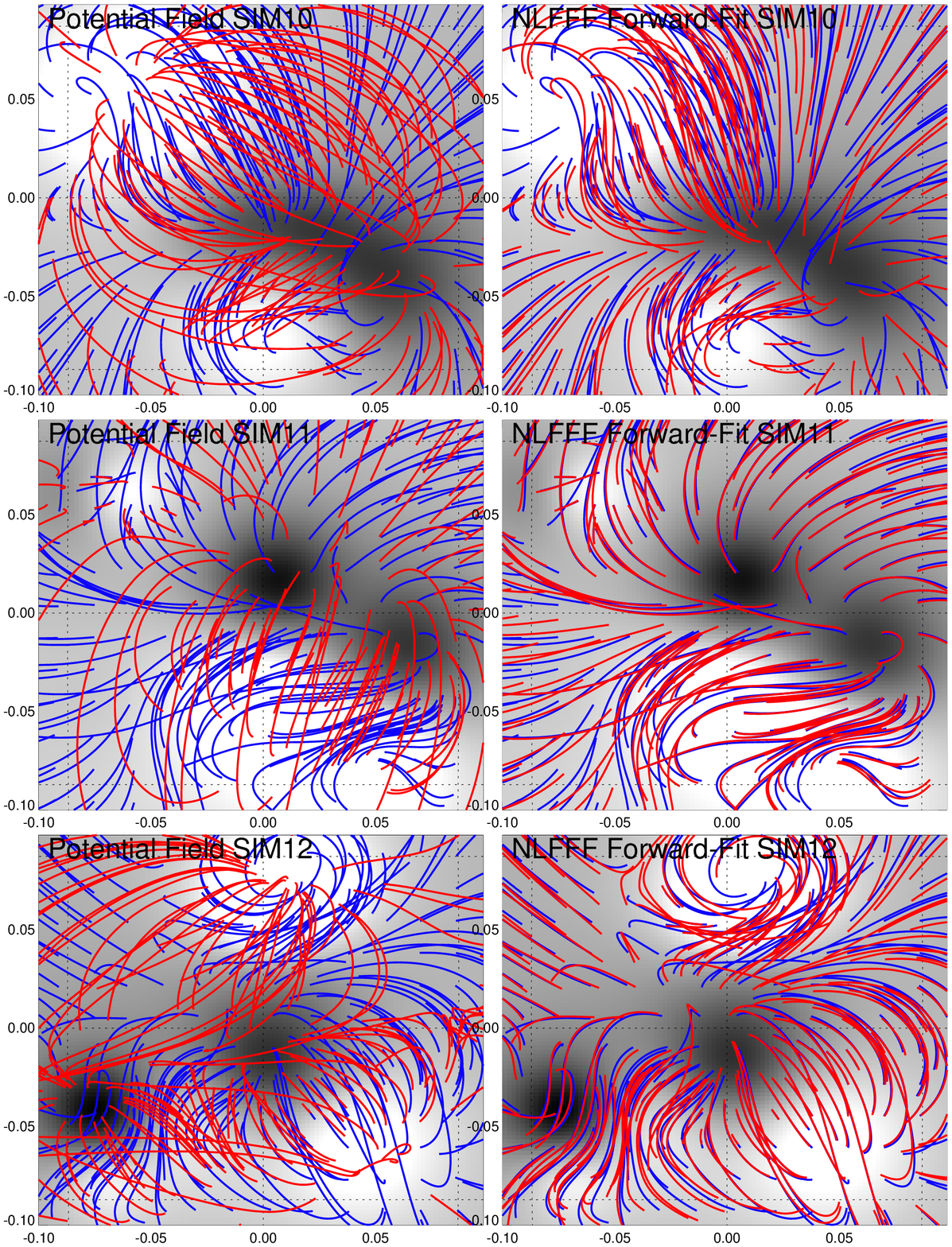}}
\caption{Forward-fitting of cases N10--N12. Representation similar
to Figure 2.}
\end{figure}

\subsection{Free Energy Estimated from Misalignment Angle with Loops}

An alternative method is to estimate the free energy in an active region 
from the misalignment angle $\mu$ between the potential field and the observed coronal
loops, or the best-fit nonpotential field. Using the coronal loops as a
proxy for the nonpotential field, this would provide a very fast method 
that requires only the computation of a potential field, supposed we have 
the 3D coordinates of coronal loops, \eg from stereoscopic triangulation. 
Since the mean azimuthal field component
is $B_\perp = B_{\rm N} \sin{(\Delta \mu)}$ and the mean radial (potential)
field component is $B_\parallel = B_{\rm N} \cos{(\Delta \mu)}$, the nonpotential 
energy ratio $q_\mu=E_{\rm N}/E_{\rm P}$ follows then directly, using the definition
of Equation (11), for the free energy,
\begin{equation}
        q_{\mu} = {E_{\rm N} \over E_{\rm P}} = {B_\perp^2 + B_\parallel^2 
	\over B_\parallel^2} = 1 + \tan^2{(\Delta \mu)} \ ,
\end{equation}
where the relative misalignment angle $\Delta \mu$ is defined as the difference
between the (median) potential $\mu_{\rm P}$ and (median) nonpotential field 
directions $\mu_{\rm N}$,
\begin{equation}
	\Delta \mu = \mu_{\rm P} - \mu_{\rm N} \ .
\end{equation}
Ideally, if the NLFFF forward-fitting code matches the coronal loops
perfectly (with $\mu_{\rm N} \approx 0$), and thus the relative misalignment angle 
of the potential field to the loops can be used, $\Delta \mu \approx
\mu_{\rm P}$. However, in reality there is always a significant difference
between the best-fit NLFFF solution and the observed loop data, either
due to stereoscopic measurement errors or due to an additional field
misalignment that cannot be described with the particular parameterization
of our NLFFF approximation.  

\section{	Numerical Tests	and Results 			}

We perform now tests of the calculation of the free magnetic energy using
the definitions given in Equations (12) and (13) for four different
datasets, using (1) simulated data produced by our analytical NLFFF 
approximation, (2) the NLFFF solution of Low and Lou (1990) with 
a known exact analytical solution, (3) active region NOAA 10930 during 
an X3.4 flare modeled by Schrijver \etal (2008) and Malanushenko \etal
(2012), and (4) stereoscopic data from four observed active regions.

\subsection{Tests with Simulated Data (P1-N12)}

We simulated six cases of potential fields (shown in Figure 3 of
Paper II), and six cases with
nonpotential fields (Figures 2 and 3), which are similar
to the cases N7 to N12 in Paper II, except that we reduced the
amount of twist by a factor of five, in order to make them more
comparable with observations of real active regions, which have
free energies of order $\lapprox 10\%$. We are using the 
parameterization of our analytical NLFFF approximation, described
in Paper I, containing 1, 2, 4, and 10 magnetic 
source components with variable vertical twist, labeled as cases 
P1--P6 and N7--N12. We integrate the nonpotential magnetic energy
$E_{\rm N}$ in a box that covers the displayed field-of-view centered at the
center of the solar disk and has a height range of $h=0.15$ solar radii 
above the photosphere. The values of the potential energies $E_{\rm P}$ and
the ratio of the nonpotential to the potential energy of the model
$(q_{\rm model})$, or of the fit ($q_{\rm fit}=E_{\rm N}/E_{\rm P}$), are listed in Table 1. 
The free energy is $E_{\rm free}=E_{\rm N}-E_{\rm P}=E_{\rm P}(q_{\rm N}-1)$.    

\begin{table}
\setlength{\tabcolsep}{1mm}
\caption{Magnetic energy calculations are listed for 20 cases, 
including six potential-field cases (P1-P6), six nonpotential-field cases 
(N7-N12), the Low and Lou (1990) case (L1, L2), the Schrijver \etal (2008)
case (S1, S2), and four stereoscopically observed active regions (A, B, C, D),
specified by the number of magnetic charges $(N_{\rm m})$, the fraction of
magnetic energy captured by the model $(q_E)$, the number 
of simulated loops $N_{l}$, the potential energy $E_{\rm P}$ (corrected
by the factor $q_E$), the median misalignment angle $\mu_{\rm P}$ of 
the potential field, the median misalignment angle $\mu_{\rm N}$ after 
forward-fitting of the NLFFF model, 
the predicted energy ratio $q_\mu=(1.+\tan^2{(\Delta \mu)})$ 
based on the misalignment angle change $\Delta \mu=\mu_{\rm P}-\mu_{N}$,
the forward-fitted nonpotential energy ratio $q_{\rm fit}=E_{\rm N}/E_{\rm P}$, 
and the volume-integrated nonpotential 
magnetic energy ratio of the model $q_{\rm model}=E_{\rm model}/E_{\rm P}$, with
values computed by Anna Malanushenko (private communication, 2012) 
using the Low and Lou (1990) data $(^a)$,
the value of 3D-fits labeled as II.b Tables 3 and 4 of Malanushenko \etal (2012) $(^b)$,
and the value of the $Wh^+$ code with the smallest misalignment angle $\mu=24^\circ$ 
in Table 1 of DeRosa \etal (2009) $(^c)$.}
\footnotesize
\begin{tabular}{lrrrrrrrr}
\hline
Case  &$N_{\rm m}(q_{\rm E})$ & $N_{\rm loop}$& $\mu_{\rm P}$& $\mu_{\rm N}$& $E_{\rm P}$     & $q_{\mu}$& $q_{\rm fit}$& $q_{\rm model}$ \\
    &           &      &        &        & $10^{32}$ &          &               \\ 
    &           &      &        &        & (erg)     &          &               \\ 
\hline
P1 &    1(1.000) &   49 &   0.5 &   0.0 & 4.65 &   1.000 &   1.000 &   1.000 \\
P2 &    2(1.000) &  121 &   1.8 &   1.2 & 5.84 &   1.000 &   1.000 &   1.000 \\
P3 &    4(1.000) &  121 &   2.1 &   1.4 & 4.27 &   1.000 &   1.000 &   1.000 \\
P4 &   10(1.000) &  121 &   2.5 &   1.3 & 14.6 &   1.000 &   1.000 &   1.000 \\
P5 &   10(1.001) &  121 &   3.7 &   1.4 & 12.9 &   1.002 &   1.000 &   1.000 \\
P6 &   10(0.998) &  121 &   3.3 &   1.9 & 6.80 &   1.001 &   1.000 &   1.000 \\
\hline
N7 &    1(1.000) &   49 &  18.8 &   0.2 & 4.65 &   1.114 &   1.010 &   1.010 \\
N8 &    2(1.000) &  121 &  10.4 &   1.2 & 5.84 &   1.026 &   1.009 &   1.009 \\
N9 &    4(1.000) &  121 &  13.0 &   1.8 & 4.27 &   1.039 &   1.016 &   1.016 \\
N10 &   10(1.000) &  121 &  17.4 &   3.1 & 14.6 &   1.065 &   1.082 &   1.082 \\
N11 &   10(1.001) &  121 &  17.2 &   1.1 & 12.9 &   1.084 &   1.113 &   1.113 \\
N12 &   10(0.998) &  121 &  25.3 &   1.9 & 6.80 &   1.188 &   1.163 &   1.163 \\
\hline
L1 &  100(0.913) &  133 &  14.7 &   4.8 & 0.246 &   1.030 &   1.023 &   1.129$^a$ \\
L2 &  100(0.910) &   35 &   3.9 &   1.6 & 0.028 &   1.002 &   1.001 &   1.091$^a$ \\
\hline
S1: 2006/12/12 &  200(0.860) &  331 &  37.3 &  14.4 & 18.3 &   1.179 &   1.112 & $1.21\pm0.05^b$\\
S2: 2006/12/13 &  200(0.878) &   98 &  27.3 &  13.4 & 17.8 &   1.061 &   1.104 & $1.08\pm0.01^b$\\
\hline
A: 2007/04/30 &  100(0.854) &  200 &  23.8 &  21.4 & 12.7 &   1.002 &   1.006 & 1.030$^c$\\
B: 2007/05/09 &  100(0.807) &   70 &  21.2 &  17.9 & 1.08 &   1.003 &   1.023 &     \\
C: 2007/05/19 &  100(0.784) &  100 &  41.3 &  22.1 & 3.47 &   1.121 &   1.085 &     \\
D: 2007/12/11 &  100(0.810) &   87 &  22.1 &  14.7 & 7.83 &   1.017 &   1.044 &     \\
\hline
\end{tabular}
\end{table}

Table 1 demonstrates several results. First, the misalignment angle between 
the forward-fitted nonpotential field and the simulated loops is in all 12 
cases (P1,...,N12) significantly smaller ($\mu_{\rm N}=1.4^\circ\pm0.8^\circ$) than the
initial potential field misalignment angle $\mu_{\rm P}=9.6^\circ\pm8.5^\circ$,
which confirms a satisfactory convergence of the forward-fit to the simulated
target loops. Second, all ratios of nonpotential to
potential energies $E_{\rm N}/E_{\rm P}$ are larger than one, which means that the free
energy $E_{\rm free}=E_{\rm N}-E_{\rm P}$ is always positive with our definition given in
Equation (12). Third, all 12 simulated cases agree in the
nonpotential energy ratio with the simulated input model, \ie 
$q_{\rm fit}=q_{\rm model}=E_{\rm N}/E_{\rm P}$ with an accuracy of less than $10^{-5}$, 
which also confirms the perfect convergence of the forward-fit algorithm.
The nonpotential-field cases (N7--N12)
have free energies in the range of $q_{\rm fit}=E_{\rm N}/E_{\rm P}=1.010-1.163$, or up to
16\% of the potential-field energy, which are also retrieved with an
accuracy of better than $10^{-5}$.
We compare also the nonpotential energy ratios calculated from the forward-fit
($q_{\rm fit}$) and estimated from the median misalignment angles 
($q_\mu$; Equation (13)) and find an agreement of $q_\mu/q_{\rm fit}=1.02\pm0.05$
between the two methods.

\begin{figure}
\centerline{\includegraphics[width=1.0\textwidth]{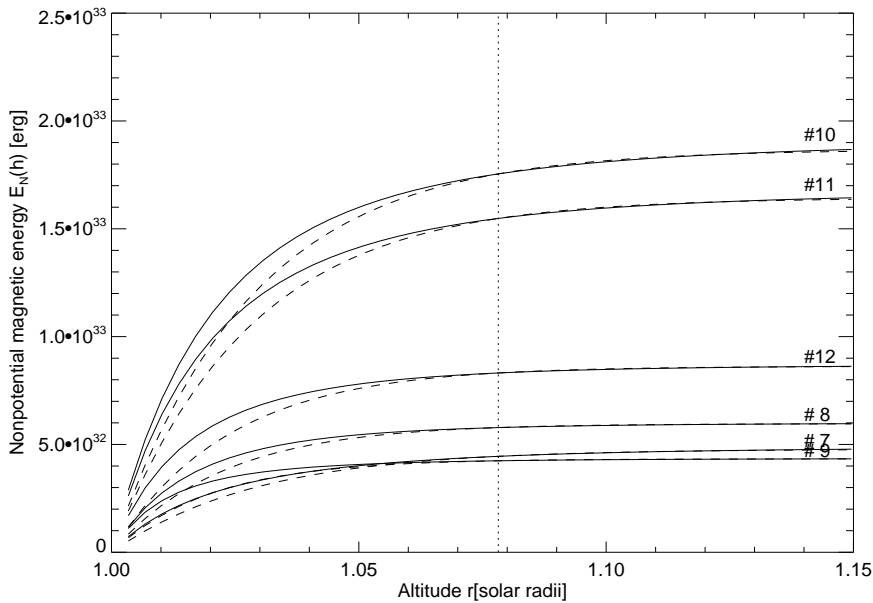}}
\caption{Height dependence of the nonpotential energy $E_{\rm N}(h)$
integrated over a volume from height $r=1.0$ to $r=r_{\rm max}$ with 
$r_{\rm max}=1.15$ solar radii for the six-nonpotential cases
(N7--N12 in Table 1). An exponential function is fitted in the upper
half height range $r>1.075$ (dotted line).}
\end{figure}

We test also the finiteness of the nonpotential energy.
In Figure 4 we plot the height-integrated total nonpotential
energies $E_{\rm N}(r)$ as a function of the height limit $r$ for the six
cases N7--N12 (Table 1) and find that each one follows approximately an
exponential height dependence (dashed curves in Figure 4),
\begin{equation}
	E_{\rm P}(r) \approx E_{\rm max} [ 1 - \exp{(-{r-1 \over \lambda})} ] \ ,
\end{equation}
which has the finite limit $E(r\mapsto \infty) = E_{\rm max}$, and thus
confirms the convergence of the code. In our calculations we generally 
use a height limit of $h_{\rm max} = 0.15$ solar radii, which corresponds
to about two density scale heights of $T_{\rm e}=1.0$ MK plasma, or four
emission measure scale heights. 

\subsection{Tests of the Low and Lou (1990) NLFFF Solution} 

An analytically exact solution of a NLFFF model was derived by 
Low and Lou (1990), described also in Malanushenko, Longcope, and
McKenzie (2009). The particular solutions we are using are defined by 
the parameters (a=0.6, n=2.0) in case L1, and (a=0.01, 
n=1.0) in the case L2, where $a$ is the Grad-Shafranov
constant, $n$ is the harmonic number of the Legendre polynomial,
and additional parameters are the depth $l$ of the source below the 
photosphere and the inclination angle $\phi$ of the axis of symmetry).
Forward-fits of our analytical NLFFF approximation to the exact NLFFF
solution of Low and Lou (1990) are shown in Figure 5 (top and middle),
where we used a computation box
of $(100 \times 100 \times 75)$, with a pixel size of $\Delta x=0.002$
solar radii and a height range of $h_{\rm max}=75 \Delta x \approx 0.15$
solar radii. The forward-fit was accomplished by using the line-of-sight
magnetogram at a planar surface ($B_z(x,y,z=1$) and a set of $N_{l}=133$
(case L1) and $N_{\rm loop}=35$ (case L2) target field lines that mimic coronal loops.
The misalignment angle is reduced from $\mu_{\rm P}=14.7^\circ$ to $\mu_{\rm N}=4.8^\circ$
for case L1, and from $\mu_{\rm P}=3.9^\circ$ to $\mu_{\rm N}=1.6^\circ$ for case L2,
so the forward-fitting reduces the misalignment by about a factor of 3.
The resulting nonpotential energy ratios are listed in Table 1, yielding 
free energies of $2.3\%$ for case L1, and $0.1\%$ for case L2, respectively. 
These free energies are significantly below the theoretical 
values calculated in Malanushenko \etal (2009), where values of 13\% and
9\% are quoted (listed under $q_{\rm model}$ in Table 1), although the 
misalignment of the forward-fitted field is quite satisfactory. 
The reasons for the mismatch in the free energy for this particular case 
is not fully understood, since the convergence behavior of our code seems
to be no problem in case of unique solutions (such as the simulated cases
N7--N12). We suspect that the parameterization of our NLFFF approximation, 
which consists of a number of buried point-like magnetic sources, is not 
adequate or suitable to represent the analytical Low and Low 
(1990) solution, which consists of extended, smooth magnetic
distributions with elliptical shapes, parameterized in terms of Legendre 
polynomials. It is conceivable that the representation of 
Legendre polynomials by (spherically symmetric) point sources
leads to clustering of closely-spaced point sources with cancelling
of the nonpotential (azimuthally twisted) field components. 

\subsection{Tests of the case of Schrijver \etal (2008)}

The magnetic field of active region NOAA 10930, observed with TRACE,
{\it Hinode}/XRT, and {\it Hinode}/SOT on 2006 December 12, 20:30 UT before a 
GOES-class X3.4 flare (case S1), and on 2006 December 13, 03:40 UT  
after the flare (case S2), has been extensively modeled with NLFFF 
codes (Schrijver \etal 2008; Malanushenko \etal 2012). We show a 
forward-fit of our NLFFF approximation in Figure 5 (bottom panels) 
to a set of loops (\ie closed field lines that are randomly chosen 
from a NLFFF solution computed by Malanushenko \etal 2012). 
The accuracy of the forward-fitting depends mostly on the number of
magnetic field components $N_{\rm m}$, but generally reaches asymptotically
a flat plateau for $N_{\rm m} \gapprox 100$ (Aschwanden \etal 2012; Figure 10).
For this magnetically very complex active region we needed $N_{\rm m}=200$ 
magnetic source components to reach the plateau, while $N_{\rm m} \lapprox 100$
was sufficient for all other cases. Here, the median
misalignment angle of $\mu_{\rm P}=37.3^\circ$ for the potential field was reduced
by about a factor of 2.6 to $\mu_{\rm N}=14.4^\circ$ for the best-fit 
nonpotential NLFFF solution (case S1 in Table 1). We measure a
potential-field energy of $E_{\rm P}=1.83 \times 10^{33}$ erg before the flare (S1),
and $E_{\rm P}=1.78 \times 10^{33}$ erg after the flare (S2), so a small difference
of $\approx 2.6\%$. For the nonpotential magnetic energy ratio we measure 
$q_{\rm fit}=E_{\rm N}/E_{\rm P}=1.112$ before the flare and $q_{\rm fit}=1.104$ after the flare, 
and similar values with the misalignment method, \ie $q_\mu=1.179$ and 
$q_\mu=1.061$. Thus the total nonpotential energy decreases by 
$\Delta E_{\rm N}=0.7\times 10^{32}$ erg ($\approx 5\%$ according to the 
forward-fit method), or by $\Delta E_{\rm N}=2.7 \times 10^{32}$ erg 
($\approx 13\%$ (according to the misalignment method). 

From the same observing times the free energy ratio was
measured with 14 different NLFFF codes in Schrijver \etal (2008; Table 1
therein), which yield energy ratios of $E_{\rm N}/E_{\rm P, free}=1.05\pm0.05$ before 
the flare, and $E_{\rm N}/E_{\rm P, free}=1.16\pm0.14$ after the flare, if we
average all methods with equal weight. 
However, the most reliable NLFFF method among them, according to a
quality assessment by visual inspection of five magnetic features 
seems to be the $Wh^+_{pp}$ NLFFF code, which yields an energy ratio of
$E_{\rm N}/E_{\rm P,pre}=1.32$ before the flare and $E_{\rm N}/E_{\rm P,pre}=1.19$ after the
flare, so a decrease of 13\% in the free energy, corresponding to a drop
of $\Delta E_{\rm free} \approx 3 \times 10^{32}$ erg in free energy, similar to our
measurement with the misalignment method ($E_{\rm free}=2.7 \times 10^{32}$ erg
or 13\%). Similarly, Malanushenko \etal (2012; cases with 3D fits labeled
as II.b in Table 1 therein) calculates energy ratios
of $E_{\rm N}/E_{\rm P}=1.21\pm0.05$ before the flare and $E_{\rm N}/E_{\rm P}=1.08\pm0.01$ 
after the flare, corresponding to a drop of 13\% in the free energy. 

Thus all three studies agree with a drop in free energy, by an amount of
$5\%-13\%$ according to our two methods, versus $\approx 13\%$ for the
most reliable NLFFF codes, while lower free energy values 
and larger misalignment angles result for the other NLFFF
codes (Table 1 in Schrijver \etal 2008). Thus our two methods appear to
be commensurable with the most reliable NLFFF codes.  

\subsection{Tests with Stereoscopic Observations}

We calculate now the free magnetic energy for four active regions 
(Table 1) that have all been observed with STEREO and were subjected 
to previous magnetic modeling, which we label as active regions A, B, C,
and D. All four active regions have been subject of potential-field 
modeling using stereoscopic data, including potential-field stretching 
(Sandman \etal 2009), buried magnetic charges (Aschwanden and Sandman, 2010), 
or buried dipoles (Sandman and Aschwanden, 2011), and nonlinear 
force-free field modeling (Aschwanden \etal 2012). 

\begin{figure}
\centerline{\includegraphics[width=1.0\textwidth]{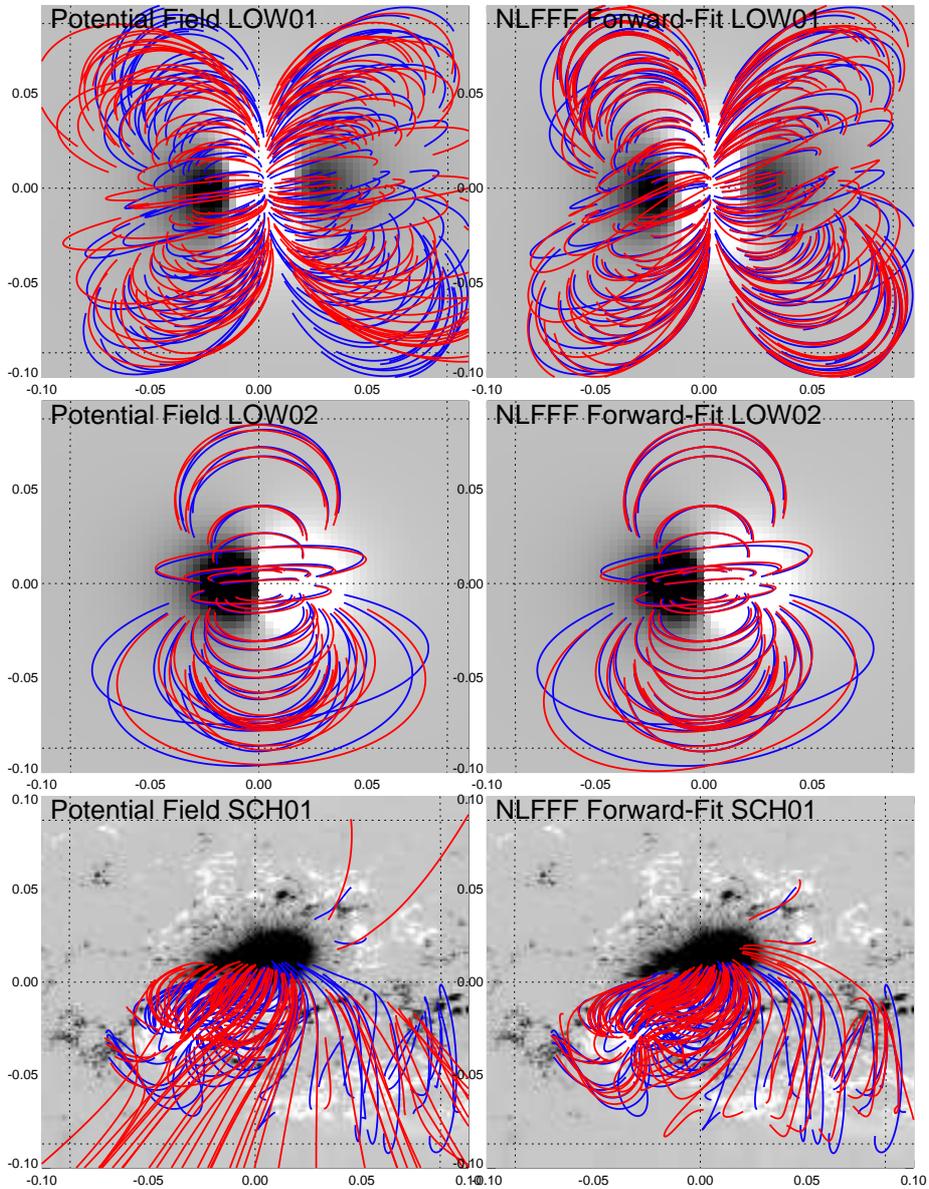}}
\caption{Forward-fitting of cases L1, L2 (Low and Lou, 1990), 
and case S1 (Schrijver \etal 2008). Representation similar
to Figure 2.}
\end{figure}

\begin{figure}
\centerline{\includegraphics[width=1.0\textwidth]{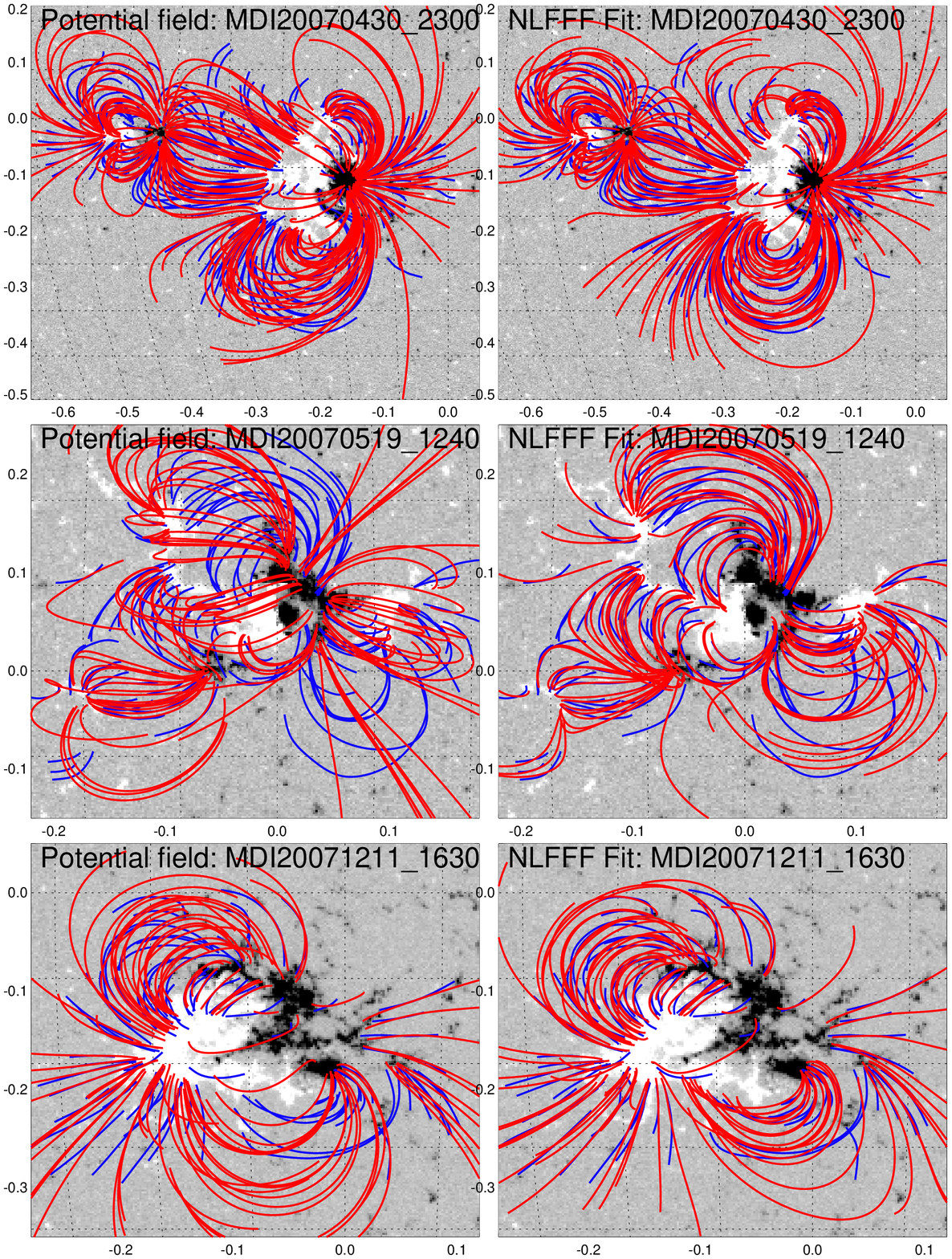}}
\caption{Forward-fitting of active region cases A (2007 April 30), 
C (2007 May 19), and D (2007 December 11). Representation similar
to Figure 2.}
\end{figure}

\medskip
\underbar{Active region NOAA 10953, 2007 April 30 (A)} has been observed
during a fla\-re energy build-up phase and a flare trigger by magnetic 
reconnection at a 3-D null point of a separatrix surface was identified
(Su \etal 2009). The magnetic modeling of this active region  
has been scrutinized with 11 different NLFFF codes, using SoHO/MDI, 
{\it Hinode}/SOT-SP, XRT, and STEREO/EUVI data (DeRosa \etal 2009; Su \etal 2009).  
For this active region (shown in Figure 6, top panels)
we measure a potential field energy of $E_{\rm P}=1.27 \times 10^{33}$ erg
on 2007 April 30, while Su \etal (2009) measure $E_{\rm P}=0.96 \times 10^{33}$ erg
on 2007 May 2, which agrees within $\approx 30\%$, over a time difference of two days.
For the nonpotential energy ratio we find $E_{\rm N}/E_{\rm P}=1.006$ on 2007 April 30,
while Su \etal (2009) find $\approx 1.1$ on 2007 May 2, in the central 
core of the active region, a few hours before a GOES-class B3.8 and C8.5
flare. Their higher value could thus be attributed to flaring
activity. Extensive NLFFF modeling was carried out for 2007 April 30
at 22:24 UT, using {\it Hinode} vector magnetograph data (DeRosa \etal 2009).
The free energies obtained from 11 different NLFFF codes scatter in the
range of $E_{\rm N}/E_{\rm P}=0.87-1.24$, with a value of 1.03 for the $Wh^+$ code with
the smallest misalignment angle of ($\mu=24^\circ$) with respect 
to the stereoscopically triangulated loops.
Including the uncertainties of the boundary conditions in the NLFFF code,
a self-consistent NLFFF solution with a nonpotential energy ratio of
$E_{\rm N}/E_{\rm P} \approx 1.08$ was obtained, with a potential-field energy of
$E_{\rm P}=0.84 \times 10^{33}$ erg (Wheatland and Leka, 2011).

Part of the discrepancy could possibly be explained by the different field-of-view,
since the {\it Hinode} field-of-view used in DeRosa \etal (2009) covers only
the central part of the active region, while our field-of-view encompasses 
the entire active region. Thus we calculated the magnetic
energy in the same {\it Hinode} field-of-view as used in DeRosa 
\etal (2009) but find almost identical energy ratios.
Our lower value than obtained with the other NLFFF codes could also 
be attributed to an underestimation of the twist (and thus nonpotential
energy) in the core of the active region, where stereoscopic loop
triangulation is very sparse due to confusion of loops with ``moss''
background. If this is true, we generally expect that the avoidance of
twisted core structures leads to a stereoscopic undersampling
bias, resulting into lower estimates of the free magnetic energy. 
We also have to keep in mind that the free energy ratio 
for this active region is the lowest among the four active regions, 
and thus has the largest relative uncertainty. 

\medskip
\underbar{Active region NOAA 10953, 2007 May 9 (B)} was subject to 
the first 3D reconstruction with STEREO (Aschwanden \etal 2008a), 
stereoscopic electron density and temperature measurements 
(Aschwanden \etal 2008b), and instant stereoscopic tomography and 
DEM modeling (Aschwanden \etal 2009). This active region exhibits
the simplest bipolar magnetic configuration among all four analyzed
active regions and we find a nonpotential energy ratio of
$E_{\rm N}/E_{\rm P}=1.023$ (Table 3), \ie a free energy ratio of $\approx 2.3\%$.

\medskip
\underbar{Active region NOAA 10953, 2007 May 19 (C)} has exhibited 
multiple filament eruptions in the complex and highly nonpotential 
magnetic configuration during 2007 May 19  
(Li \etal 2008; Liewer \etal 2009; Hara \etal 2011). 
Some 22 GOES B-class and 2 GOES C-class flares were detected during
the observing period (Li \etal 2008). A filament eruption, accompanied
by a B9.5 flare, coronal dimming, and and EUV wave was observed and
traced with 3D reconstruction after 2007 May 19, 13:00 UT (Liewer \etal
2009). The associated EUV dimming and EUV wave caused by the filament
eruption was also analyzed (Attrill \etal 2009).
Plasma motion and heating up to $T_{\rm e}=9$ MK was observed for
the same flare around 13:00 UT (Hara \etal 2011). 
For this active region, which we analyzed at 12:47 UT shortly before
the flare (shown in Figure 6, middle panels), 
we found the largest amount of free energy (9\%), \ie 
a nonpotential energy ratio of $E_{\rm N}/E_{\rm P}=1.085$ (Table 1), which clearly 
is associated with the filament eruption and flaring activity after 13:00 UT. 

\medskip
\underbar{Active region NOAA 10978, 2007 December 11 (D)} appears also to
have a dominant bipolar structure (shown in Figure 6, bottom panels), 
but some apparent currents along
the central neutral line have been modeled with a flux-insertion method
(Alex Engell and van Ballegooijen; private communication 2012). 
For this active region we found a moderate amount of free energy (4\%), 
\ie a nonpotential energy ratio of $E_{\rm N}/E_{\rm P}=1.044$ (Table 1), which 
is likely to be stored in the flux rope or filament above the central
neutral line. 

\begin{figure}
\centerline{\includegraphics[width=1.0\textwidth]{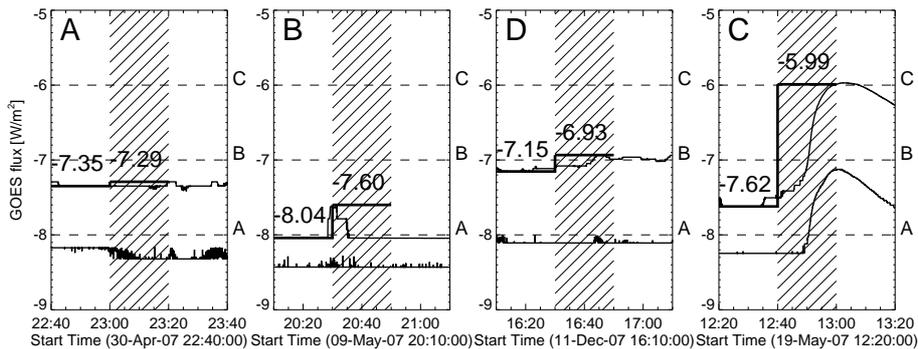}}
\caption{GOES soft X-ray light curves of the 0.5-4 \ang\ (upper
curve) and 1-8 \ang\ channel (lower curve) during the time of
stereoscopic triangulation and magnetic modeling of the four
analyzed active regions. The flare peak and preflare background
levels are indicated with a step function (thick solid lines).
The four active regions are in order of increasing GOES flux
(adapted from Aschwanden and Sandman, 2010).}
\end{figure}

\begin{figure}
\centerline{\includegraphics[width=0.8\textwidth]{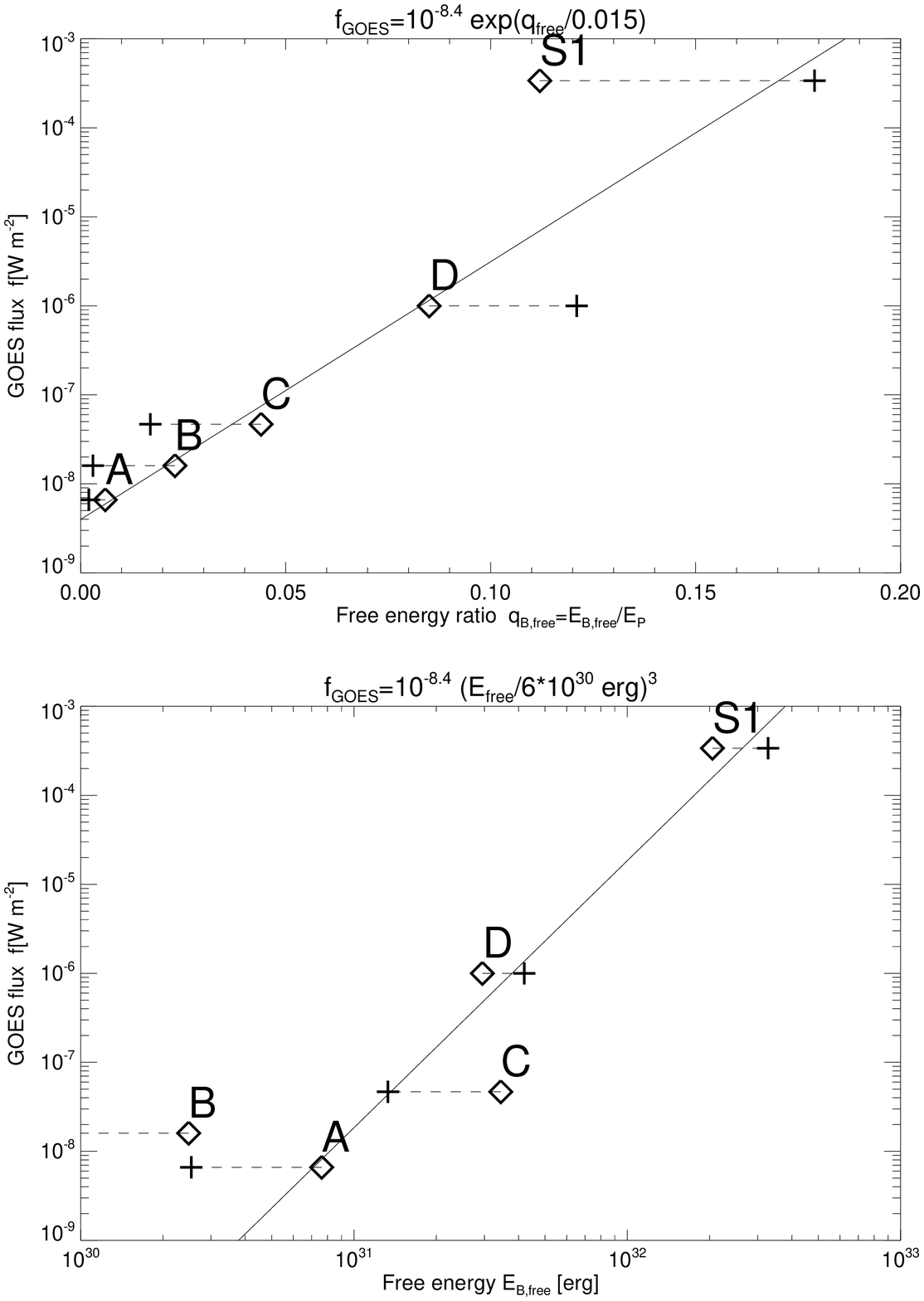}}
\caption{Correlation between the free energy ratio 
$q_{\rm free}=E_{\rm free}/E_{\rm P}$ (top panel) or the free energy $E_{\rm free}$
versus the logarithmic (preflare background-subtracted) GOES 0.5-4 
\ang\ flux $f_{\rm GOES}$ for active regions A, B, C, and D, are shown,
determined with the forward-fit method (diamonds) and with the
misalignment method (crosses). Linear regression fits are indicated
between linear or logarithmic values are indicated, yielding an 
exponential function (top panel) or a powerlaw function (bottom
panel).}
\end{figure}

\subsection{Correlation of Free Energies with Flare Fluxes}

It was noted in earlier work that the level of flaring activity in
these four active regions is positively correlated with the
nonpotentiality of the magnetic field, evaluated by the average
misalignment angle between a magnetic potential-field model and the
observed 3D loop coordinates. This correlation
was interpreted in terms of a relationship between
electric currents and plasma heating (Aschwanden and Sandman, 2010). 
Since the misalignment angle is a measure of the magnetic nonpotentiality, 
we expect that there should also be a correlation between the flaring 
activity level and the free magnetic energy in a flaring active region. 
The GOES 0.5-4 and 1-8 \ang\ light curves are shown in Figure 7.
We plot the (preflare background-subtracted) 
GOES 0.5-4 \ang\ fluxes $f_{\rm GOES}$ of the largest flare that occurred 
during the observing period of an active region versus the free energy 
ratio $q_{B,\rm free}=E_{B,\rm free}/E_{\rm P}$ in Figure 8 (top panel). 
The preflare background-subtracted GOES fluxes $F_{\rm GOES}$
show a correlation that can be fitted with an exponential function,
\begin{equation}
	\left( {f_{\rm GOES} \over f_0} \right) \approx 
	\exp{\left({ q_{\rm free} \over q_0} \right)} \ .
\end{equation}
with the constants $q_0=0.015$ and $f_0=10^{-8.4}$ (W m$^{-2})$.
This result implies that the magnitude of the flare (measured here
with the GOES soft X-ray flux) is directly related to the
free magnetic energy stored in the active region before the flare.
If we include the X3.4 GOES flare of AR 10930 (Figure 8, case S1),
we see that the observed free energy ratio $q_{\rm fit}=E_{\rm N}/E_{\rm P}-1 \approx 0.112$ 
lies along the same trend of the extrapolated exponential function.

Also the free energy $E_{\rm free}$, an absolute measure of the nonpotential 
energy, shows a correlation with the observed GOES flux, approximately
following a powerlaw (Figure 8, bottom),
\begin{equation}
	\left( {f_{\rm GOES} \over f_0} \right) \approx 
	\left( { E_{\rm free} \over E_0} \right)^3 \ .
\end{equation}
with the constants $f_0=10^{-8.4}$ (W m$^{-2})$ and $E_0=6 \times 10^{30}$ erg.
Only the active region (B) with the lowest free energy is an outlier to this 
powerlaw relationship. 

\section{Discussion} 

\subsection{Accuracy of the Free Energy}

Compared with the other (numerical) NLFFF codes, our analytical NLFFF 
code has the following
advantages: (1) computational speed that allows fast forward-fitting to
observed coronal data; (2) simplicity of explicit analytical formulation;
(3) spherical geometry of solar surface
is fully implemented; and (4) the free energy meets the five criteria of 
positivity, additivity of energy (or orthogonality of magnetic field
components), positive scaling with force-free parameter 
$\alpha$, potential field limit for small $\alpha$'s, and finiteness
of nonpotential energy. In this study we calculated the free magnetic
energy for four different test datasets and aim to validate the
accuracy. 

The first test dataset consists of simulated loops, which
are analytically defined by the same parameterization as the magnetic
field lines that are forward-fitted to the simulated data. We find
that the total nonpotential magnetic energies $E_{\rm N}$ are retrieved 
with an accuracy of $\approx 10^{-5}$ for these cases, and the free 
energy ratios $E_{\rm N}/E_{\rm P}$ are retrieved with an accuracy of 
$\approx 10^{-3}$. The high accuracy just confirms the convergence
behavior of our code in the case of a unique solution, as it is
the case when the forward-fitting model has the same parameterization
as the fitted simulation dataset.

The second test dataset consists of an analytical NLFFF solution,
which has a completely different parameterization in terms of the 
Green's function applied to constant-$\alpha$ point sources 
(Chiu and Hilton, 1977; Low and Lou, 1990; Lothian and Browning, 1995; 
Malanushenko \etal 2009, 2011, 2012). The particular case described
in Low and Lou (1990) consists of three smooth elliptical magnetic
field concentrations and contains an amount of 13\% free energy,
while we recovered only 2\% with the forward-fit.
The poor performance is possibly 
due to the particular topology of the Low and Lou (1990) model,
which has extended (elliptical) magnetic field regions parameterized
with smooth Legendre polynomials, which cannot be fitted uniquely with 
spherical point sources. However, because this case is untypical for solar 
observations, which appear to consist of many small-scale flux
concentrations that can be easier fitted with point sources, the
accuracy of free energies may be much better for real solar data.

The third data set consists of an observed flaring active region
with a large X3.4 flare, for which a NLFFF solution was calculated. 
For the free energy we obtain a value of order 11\%-18\% before the
flare, which approximately agrees with the most reliable 
NLFFF solution (16\%-32\%) computed in Schrijver \etal (2008) and
Malanushenko \etal (2012). Also for the sign of the change in
nonpotential magnetic energy during this X3.4 flare we obtain
the same sign and a similar value for the decrease of free energy 
($5\%-13\%$) as the values computed in Schrijver \etal (2008) and
Malanushenko \etal (2012), $\approx13\%$.

The fourth test dataset consists of four solar active regions, for 
which the 3D coordinates of coronal loops could be stereoscopically
triangulated. For these four cases we obtained free energies of
0.6\%, 2.3\%, 4.4\%, and 8.5\%. Only for the first case (active
region A, 2007 April 30) we have comparisons with other NLFFF codes. 
The most reliable NLFFF code that exhibits the smallest misalignment angle 
with stereoscopic loops, yield 3\% free energy, which is close 
to our result of $E_{\rm free}/E_{\rm P}=0.6\%$. The free energy
obtained from all 11 NLFFF solutions yields a much larger scatter,
\ie $E_{\rm free}/E_{\rm P}=10\%\pm12\%$ (Table 1 in DeRosa \etal 2009). 

\subsection{Scaling Law between Magnetic Energy and Flare Soft X-ray Flux}

We found a correlation between the free magnetic energy and the GOES flux
of the largest flare that occurred during the observing period (Equation 16).
Since our nonpotential field solution is parameterized by a vertical
twist of magnetic charges, the free energy is directly proportional
to the magnetic energy associated with the azimuthal field component
$B_{\varphi}$, and thus approximately proportional to the squared force-free 
$\alpha$ parameter or the number $N_{\rm twist}$ of twists per length unit,
(Equations (10) and (6)),
\begin{equation}
	q_{\rm free}({\bf x}) = {dE_{\rm N}({\bf x}) \over dE_{\rm P}({\bf x})}-1 = 
	{B_\varphi^2 \over B_r^2} = (b r \sin{\theta})^2 
	\propto N_{\rm twist}^2 \propto \alpha^2 \ .
\end{equation}
On the other side, the stress-induced heating rate $E_{\rm H}$ by Ohmic dissipation 
(or Joule dissipation) is proportional to the square of the currents,
(\eg van Ballegooijen, 1986), which is also proportional to the square 
of $\alpha$ (with ${\bf j}/4\pi = \nabla \times {\bf B} = \alpha {\bf B}$), 
and thus to the free energy, 
\begin{equation}
	E_{\rm H} = {j^2 \over \sigma} \propto {\alpha^2 B^2 \over 4\pi\sigma} \ ,
\end{equation}
where $\sigma$ is the classical conductivity. Thus we can express
Equation (16) in terms of twists, currents, or heating rates,
\begin{equation}
	F_{\rm SXR} \propto \exp{(q_{\rm free})} 
		\propto \exp{(N_{\rm twist}^2)} 
		\propto \exp{(\alpha^2)}
		\propto \exp{(j^2/B^2)} 
		\propto \exp{(E_{\rm H}/B^2)} 
		\ .
\end{equation}
Our empirical finding suggests that the energy radiated from the heated
plamsa is not just proportional to the heating energy, but 
``exponentiated'' by the current density $j$, the force-free parameter
$\alpha$, the number of twists $N_{\rm twist}$, or the heating rate $E_{\rm H}$ of
Joule dissipation. This implies a highly nonlinear mechanism that
converts vertical twist into thermal radiation via dissipated currents,
as envisioned in stress-induced reconnection (Sturrock and Uchida, 1981;
Parker, 1983). Our linear regression fit of the free energy versus
the GOES flux yields a lower limit of $F_{\rm GOES}=10^{-8.4}$, which
corresponds to a GOES-class Z4-event, about the magnitude of the
smallest detectable nanoflare. The largest flare among our analyzed 
cases is an X3.4 GOES-class flare. Apparently, the exponential 
relationship between soft X-ray flux and free energy discovered here 
holds even approximately for large GOES X-class flares (Figure 8).
The nonlinearity is also reflected by the fact that the thermal energy
of the soft X-ray flux is not just proportional to the free magnetic
energy, but rather exhibits a highly nonlinear powerlaw function with
a powerlaw index of $\approx 3$ (Equation (17)).

\section{Conclusions}

We calculated the total free magnetic energy contained in a coronal volume
encompassing an active region for four different kinds of datasets:
(1) simulated data, (2) data created from an analytical NLFFF
solution by Low and Lou (1990), (3) a flaring active region,
and (4) four active regions observed
with STEREO and SOHO/MDI. The free magnetic energy $E_{\rm free}=E_{\rm N}-E_{\rm P}$
is defined by the difference of the nonpotential ($E_{\rm N}$) and the potential 
magnetic field energy ($E_{\rm P}$). The nonpotential magnetic field 
${\bf B}_{\rm N}({\bf x})$
is defined by an analytical approximation of a NLFFF solution that is
parameterized by buried magnetic charges with vertical twists (derived
in Paper I). The numerical code that performs fast forward-fitting of
magnetic field lines to coronal 3D constraints, such as stereoscopically
triangulated loops, is described in Paper II, along with the simulated data.
Our findings are as follows:

\begin{enumerate}
\item{A first method to calculate the free energy results from 
	forward-fitting of our analytical NLFFF approximation by
	associating the perpendicular magnetic field component $B_{\perp}$
	with the free energy, $dE_{\rm free}=B_{\perp}^2/8\pi$, while the
	parallel component $B_\parallel$ is associated with the potential field
	energy $dE_{\rm P}=B_\parallel^2/8\pi$. This definition of the free energy 
	fulfills the conditions of (i) positivity of free energy,
	(ii) additivity of energies, $E_{\rm N} = E_{\rm P} + E_{\rm free}$
	and orthogonality $B_\perp({\bf x}) \perp B_{\rm P}({\bf x})$,
	(iii) a positive scaling with the force-free parameter, $dE_{\rm free}
	\propto \alpha^2$, (iv) the potential-field limit, $E_{\rm N}(\alpha
	\mapsto 0) = E_{\rm P}$, and (v) the finiteness of the nonpotential
	energy with height $h$, $E_{\rm N}(h\mapsto \infty)=E_{\rm max}$.}
\item{A second method to estimate the free energy can be obtained
	from the mean misalignment angle $\Delta \mu = \mu_{\rm P}-\mu_{\rm N}$ 
	between a potential
	and a nonpotential field (or stereoscopically triangulated
	coronal loops). The free energy ratio is then $E_{\rm free}/E_{\rm P}
	\approx \tan^2{(\Delta \mu)}$. We find that the uncertainty
	of this method amounts to $\approx \pm2\%$ for the 
	nonpotential magnetic energy.}
\item{Calculating the free energies for the simulated data we find
	a high fidelity of order $10^{-5}$ in retrieving the free
	energy, which is due to the fact that the simulated data
	have the same parameterization as the forward-fitting method,
	constraining a single best-fit solution.}
\item{Calculating the free energy for the Low and Lou (1990) analytical
	case, our NLFFF code finds a significantly lower value than
	theoretically calculated, probably because of the special 
	morphopology (parameterized with smooth Legendre functions),
	which cannot adequately be fitted with our NLFFF code that is
	designed for spherical magnetic point sources, as found in
	solar magnetograms.}
\item{Calculating the free energy for observed active regions 
	constrained by the 3D coordinates of stereoscopically
	triangulated coronal loops, we find free energy ratios of
	$q_{\rm free}=E_{\rm free}/E_{\rm P}\approx 1\%-10\%$. The uncertainty of
	the free energy determined with our forward-fitted NLFFF approximation
	appears to be at least as good as the uncertainty among
	other (standard NLFFF extrapolation) codes.}
\item{We find also a correlation between the free magnetic energy
	$E_{\rm free}$ and the GOES flux of the largest flare that occurred
	during the observing period, which can be quantified by an
	exponential relationship, $F_{\rm GOES} \propto \exp{(q_{\rm free})}$,
	implying an exponentiation of the dissipated currents.}
\end{enumerate}

In summary, this study demonstrates that the free energy in active
regions can be calculated and predicted with our analytical NLFFF approximation 
with an accuracy that is commensurable with other standard NLFFF codes.
Our code has the additional advantages of computational speed for forward-fitting
of coronal data, correct treatment of the curved solar surface, positivity, 
and finiteness of free energy.
In addition, forward-fitting of our NLFFF approximation achieves a significantly 
smaller misalignment angle with respect to the observed coronal loops
($\mu \approx 2^\circ-22^\circ$), compared with the results of other NLFFF codes
($\mu=20^\circ-44^\circ$; DeRosa \etal 2009). The most limiting drawback of our
method is the availability of stereoscopic data with suitable 
spacecraft separation angle (which was most favorable in 2007, the
first year of the STEREO mission). In future work we attempt to
circumvent the 3D geometry of coronal loops by using only the 2D projections
of coronal loops, which can (manually or automatically) be traced from
loop-rich EUV or soft X-ray images and do not require stereoscopic
data at all. 

\acknowledgements
The author appreciates the provided data and helpful discussions 
with Anna Malanushenko and Marc L. DeRosa. Part of the work was supported by
NASA contract NNG 04EA00C of the SDO/AIA instrument and
the NASA STEREO mission under NRL contract N00173-02-C-2035.

\section*{References} 

\def\ref#1{\par\noindent\hangindent1cm {#1}}

\small
\ref{Aschwanden, M.J.: 2004, {\sl Physics of the Solar Corona. An Introduction},
        Praxis Publishing Co., Chichester UK, and Springer, Berlin, 
	Section 5.3.}
\ref{Aschwanden, M.J., W\"ulser, J.P., Nitta, N.V., Lemen, J.R.:
	2008a, \apj {\bf 679}, 827.}
\ref{Aschwanden, M.J., Nitta, N.V., W\"ulser, J.P., Lemen, J.R.:
	2008b, \apj {\bf 680}, 1477.}
\ref{Aschwanden, M.J., W\"ulser, J.P., Nitta, N.V., Lemen, J.R., 
	Sandman, A.W.: 2009, \apj {\bf 695}, 12.}
\ref{Aschwanden, M.J., Sandman, A.W.: 2010, Astronom. J. {\bf 140}, 723.}
\ref{Aschwanden, M.J., W\"ulser, J.P., Nitta, N.V., Lemen, J.R.,
	DeRosa, M.L., Malanu\-shen\-ko, A: 2012, \apj {\bf 756}, 124.}
\ref{Aschwanden, M.J.: 2012, \sp (online-first), 
	DOI: 10.1007/s11207-012-0069-7, (Paper I).}
\ref{Aschwanden, M.J., Malanushenko, A.: 2012, \sp (online-first), 
	DOI: 10.1007 /s11207-012-0070-1, (Paper II).}
\ref{Attrill, G.D.R., Engell, A.J., Wills-Davey, M.J., Grigis, P.,
	Testa, P.: 2009, \apj {\bf 704}, 1296.}
\ref{Chiu, Y.T., Hilton, H.H.: 1977, \apj {\bf 212}, 873.}
\ref{Choe, G.S., Cheng, C.Z.: 2002, \apjl {\bf 574}, L179.}
\ref{DeRosa, M.L., Schrijver, C.J., Barnes, G., Leka, K.D., Lites, B.W.,
        Aschwanden, M.J., \etal: 2009, \apj {\bf 696}, 1780.}
\ref{DeVore, C.R., Antiochos, S.K.: 2005, \apj {\bf 628}, 1031.}
\ref{Falconer, D., Moore, R., Gary, G.A.: 2006, \apj {\bf 644}, 1258.}
\ref{Falconer, D., Barghouty, A.F., Khazanov, I., Moore, R.:
	2011, Space Weather {\bf 9}(4), CiteID S04003.}
\ref{Fang, F., Manchester, W.IV., Abbett, W.P., van der Holst, B:
	2012, \apj {\bf 754} 15.}
\ref{Hara, H., Watanabe, T., Harra, L.K., Culhane, J.L., Young, P.R.:
	2011, \apj {\bf 741}, 107.}
\ref{Jing, J., Chen, P.F., Wiegelmann, T., Xu,Y., Park, S.H., Wang H.:
	2009, \apj {\bf 696}, 84.}
\ref{Jing, J., Tan,C., Yuan, Y., Wang, B., Wiegelmann, T., Xu,Y., Wang H.:
	2010, \apj 713, 440.}
\ref{Kusano, K., Maeshiro, T., Yokoyama, T., Sakurai,T.: 
	2002, \apj {\bf 577}, 501.}
\ref{Li, Y., Lynch, B.J., Stenborg, G., Luhmann, J.G., Huttunen, K.E.J.,
	Welsch, B.T., Liewer, P.C., Vourlidas, V.: 
	2008, \apjl {\bf 681}, L37.}
\ref{Liewer, P.C., De Jong, E.M., Hall, J.R., Howard, R.A., Thompson, W.T.,
	Culhane, J.L., Bone, L., van Driel-Gesztelyi, L.: 
	2009, \sp {\bf 256}, 57.}
\ref{Lothian, R.M., Browning, P.K.: 1995, \sp {\bf 161}, 289.}
\ref{Low, B.C., Lou, Y.Q.: 1990, \apj {\bf 408}, 689.}
\ref{Malanushenko, A., Longcope, D.W., McKenzie, D.E.: 2009,
        \apj {\bf 707}, 1044.}
\ref{Malanushenko, A., Yusuf, M.H., Longcope, D.W.: 2011,
        \apj {\bf 736}, 97.}
\ref{Malanushenko, A., Schrijver, C.J., DeRosa, M.L., Wheatland, M.S.,
	Gilchrist, S.A.: 2012, \apj {\bf 756}, 153.}
\ref{Metcalf, T.R., Jiao, L., Uitenbroek, H., McClymont, A.N., Canfield, R.C.:
	1995, \apj {\bf 439}, 474.}
\ref{Metcalf, T.R., Leka, K.D., Mickey, D.L.: 2005, \apjl {\bf 623}, L53.}
\ref{Parker, E.N.: 1983, \apj {\bf 264}, 642.}
\ref{R\'egnier, S., Priest, E.R.: 2007, \aap {\bf 669}, L53.}
\ref{R\'egnier, S.: 2009, \aap {\bf 497}, 17.}
\ref{R\'egnier, S.: 2012, \sp {\bf 277}, 131.}
\ref{Sandman, A.W., Aschwanden, M.J., DeRosa, M.L., W\"ulser, J.P.,
	Alexander D.: 2009, \sp {\bf 259}, 1.}
\ref{Sandman, A.W., Aschwanden, M.J.: 2011: \sp {\bf 270}, 503.}
\ref{Schrijver, C.J., DeRosa, M., Metcalf, T.R., Liu, Y., McTiernan, J., 
	Regnier, S., Valori, G., Wheatland, M.S., Wiegelmann,T.:
 	2006, \sp {\bf 235}, 161.}
\ref{Schrijver, C.J., DeRosa, M., Metcalf, T.R., Barnes, G., Lites, B.,
	Tarbell,T., \etal 2008, \apj {\bf 675}, 1637.}
\ref{Sturrock, P.A., Uchida, Y.: 1981, \apj {\bf 246}, 331.}
\ref{Su, Y., Van Ballegooijen, A., Lites, B.W., DeLuca, E.E., Golub, L.,
	Grigis, P.C., Huang, G.,, Ji, H.S.: 2009, \apj {\bf 691}, 105.}
\ref{Van Ballegooijen, A.A.: 1986, \apj {\bf 311}, 1001.}
\ref{Wang, H., Ewell, M.W.Jr., Zirin, H.: 1994, \apj {\bf 424}, 436.}
\ref{Wang, H.: 1997, \sp {\bf 174}, 163.}
\ref{Wang, H., Spirock, T.J., Qiu, J., Ji, H., Yurchyshyn, V., Moon, Y.J., 
	Denker, C., Goode,P.R.: 2002, \apj {\bf 576}, 497.}
\ref{Wang, H., Qiu, J., Jing, J., Spirock, T. J., Yurchyshyn, V., Abramenko, V.,
	Ji, H., Goode, P.R.: 2004, \apj {\bf 605}, 931.}
\ref{Wang, H. 2006, \apj {\bf 649}, 490.}
\ref{Wang, H., Liu,C.: 2010, \apjl {\bf 716}, L195.}
\ref{Wang, S., Liu, C., Wang, H. 2013, \apjl {\bf 757}, L5.}
\ref{Wheatland, M.S., Leka, K.D.: 2011, \apj {\bf 728}, 112.}
\ref{Wiegelmann, T.: 2004, \sp {\bf 219}, 87.}
\ref{Wiegelmann, T., Sakurai T.: 2012, Living Rev.~Solar Phys. {\bf 9}(5),
	\url{http://solarphysics.livingreviews.org/Articles/lrsp-2012-5}.}
\ref{Woltjer, L.: 1958, Proc. Ntal. Acad. Sci., {\bf 44}, 489.}

\end{article}
\end{document}